\documentclass[11pt]{article}
\usepackage{graphicx}
\usepackage[margin=1.25in]{geometry}
\usepackage[usenames,dvipsnames]{color}
\usepackage{url}
\usepackage[colorlinks = true,
            linkcolor = blue,
            urlcolor  = blue,
            citecolor = blue,
            anchorcolor = blue]{hyperref}

\newcommand\pubnumber{Transcendental Preprint }
\newcommand\pubdate{\today}

\def\Title#1{\begin{center} {\LARGE #1 } \end{center}}
\def\Author#1{\begin{center}{ \sc #1} \end{center}}

\newcommand\pubblock{\rightline{\begin{tabular}{l} \pubnumber\\
         \pubdate \end{tabular}}}
\newenvironment{Abstract}{\begin{quotation} \begin{center}
                       ABSTRACT
     \end{center}\bigskip  }{\end{quotation}}

\def\beq{\begin{equation}}
\def\eeq#1{\label{#1}\end{equation}}
\def\eeqn{\end{equation}}

\newenvironment{Eqnarray}%
   {\arraycolsep 0.14em\begin{eqnarray}}{\end{eqnarray}}
\def\beqa{\begin{Eqnarray}}
\def\eeqa#1{\label{#1}\end{Eqnarray}}
\def\eeqan{\end{Eqnarray}}

\let\bar=\overbar

\def\lsim{\mathrel{\raise.3ex\hbox{$<$\kern-.75em\lower1ex\hbox{$\sim$}}}}
\def\gsim{\mathrel{\raise.3ex\hbox{$>$\kern-.75em\lower1ex\hbox{$\sim$}}}}

\def\del{\partial}
\def\Dslash{\not{\hbox{\kern-4pt $D$}}}
\def\dslash{\not{\hbox{\kern-2pt $\del$}}}
\def\pslash{\not{\hbox{\kern-2pt $p$}}}
\def\ETmiss{\not{\hbox{\kern-4pt $E$}}_T}

\def\Dlr{\mathrel{\raise1.5ex\hbox{$\leftrightarrow$\kern-1em\lower1.5ex\hbox{$D$}}}}

\def\MSB{{\bar{M \kern -2pt S}}}
\def\msb{{\bar{\scriptsize M \kern -1pt S}}}

\def\drb{{\bar{\scriptsize D \kern -1pt R}}}

\newcommand\snowmass{\begin{center}\rule[-0.2in]{\hsize}{0.01in}\\\rule{\hsize}{0.01in}\\
\vskip 0.1in Submitted to the  Proceedings of the US Community Study\\ 
on the Future of Particle Physics (Snowmass 2021)\\ 
\rule{\hsize}{0.01in}\\\rule[+0.2in]{\hsize}{0.01in} \end{center}}

\begin{document}

\pubblock

\Title{Enabling Capabilities for Infrastructure and Workforce in Electronics and ASICs }

\bigskip 

\Author{\textbf{Marina Artuso} \\
	Syracuse University  \\
	\textbf{Carl Grace, Timon Heim} \\
	Lawrence Berkeley National Laboratory \\
	\textbf{Angelo Dragone, Ryan Herbst, Lorenzo Rota} \\
	SLAC National Accelerator Laboratory  \\
	\textbf{Gabriella Carini, Grzegorz Deptuch} \\
	Brookhaven National laboratory  \\
	\textbf{Mitch Newcomer} \\
	University of Pennsylvania  \\
	\textbf{Kevin Flood} \\
	Nalu Scientific LLC
}

\medskip

%\Address{ }

\medskip

 \begin{Abstract}
\noindent 
The impressive progress in data rate capabilities, pattern recognition, and spatial resolution of current detectors in experimental particle physics has been possible thanks to the availability of sophisticated analog and digital signal processors implemented in custom made front end application specific integrated circuits (ASICs). The performance requirements and increasingly smaller feature size chosen for these devices pose significant challenges to the designers. In order to meet these demands, we need to educate and retain a skilled work force, as well as aggregate and disseminate design tools and acquired knowledge. We propose a multifaceted initiative to ensure that these needs are met.
\end{Abstract}

\snowmass

\def\thefootnote{\fnsymbol{footnote}}
\setcounter{footnote}{0}

\section{Introduction}
Developments in the area of ASICSs and associated electronics have enabled revolutionary innovations in detector systems that facilitated measurements that gave us an unprecedented understanding of elementary particles and their interactions and of the origin of the universe \cite{Ahmed:2019sim}. The high energy physics scientific community is currently engaged in a planning exercise aiming at a vast array of future discoveries. These discoveries will be possible only if access to a shared infrastructure, shared organizational network, as well as the education and retention of experts in academia, national laboratories, and the private sector are supported in the crucial area of advanced electronics including near-detector front-end electronics implemented in application specific integrated circuits (ASICs) and field programmable gate arrays (FPGAs), used for back-end processing.
%the design, development, and construction of advanced electronics devices can be implemented.

The construction of modern particle physics detectors is a multi-year engagement, that requires a skilled and dedicated workforce. However, the demand for project-driven ASICs is uneven, and a robust R\&D program is necessary to evolve the needed capabilities. Moreover, broader access to educational modules and shared design tools are needed. 
\section{Infrastructure}
    HEP detector systems have had the benefit of many evolved layers of sophistication over   several decades. This accumulated expertise has been preserved in highly summarized form to a few archival books, summer schools, workshops, conferences and journal articles as well as internal detector notes distributed widely. 
    A more complete, detailed understanding is captured in the minds of those builders whose ideas were tested, revised and made to work  in the design and development of today's highly sophisticated readout systems. As detector systems become more complex and more granular, the time between conception and realization of a data taking system is becoming consistent with the length of a career.  As a result, the first hand direction traditionally required to efficiently evolve next generation detectors will suffer the effects of attrition 
      unless a  conscious effort is put into archiving information in a searchable form \& providing HEP specific training for incoming generations of instrumentation specialists.     
    
    The benefits of miniaturization have led to a close relationship between sensors and the front end electronics. In the most radiation intense areas of the ATLAS and CMS LHC detectors, 66000 channel pixel sensors are mated through direct bonding to custom ASICs with the complete front end readout. For example,a detailed description of the CMS Pixel detector can be found in \cite{CMSPixl:2022wul}. 
    
    Plans for future tracking detectors include placing the readout electronics on the same substrate as sensors.    
    Collider detectors depend on high collision rates and as a result operate in extreme radiation environments. Multi-kiloton liquid noble element detectors operate sophisticated readout electronics instrumentation immersed at temperatures hundreds of degrees lower than commercial vendors qualify the performance of their electronics at.     
   These extreme environments have  made it mandatory to understand the performance of commercially available electronics especially custom integrated circuits  in extreme environments far from the manufactures intended use.  These non commercial research uses require careful measurements and the development of device models for basic design elements over the range of fabrication extremes provided by fabrication facilities.   These teams of specialists who delve into the physics of the devices and understand the required operating conditions for the commercial processes and in turn develop libraries of parts adapted for the intended environmental extreme.   To make progress, this  research effort needs to be disseminated among  the community of investigators who learn as they develop new designs. This uneaven basic knowledge makes accurate predictions of development schedules difficult and leads to varied results for different groups. 
   
   A notable recent example was the realization that several complex ASIC designs nearing completion for the High Luminosity LHC detector sub-systems were behind in their scheduled completion dates.  On examination it was recognized that although the complexity of custom IC's has dramatically increased as well as the number of specialized design tools the size of some design teams remained the same. 
 Details of the new capabilities of HL-LHC designs include the fact that recent advances in ASIC technology have lower sensitivity to ionizing radiation ($>$ 100 MRad) enabling a higher density of gates and operational speed resulting in increased data transfer rates. At the same time the size of sensing elements has decreased  resulting in a requirement for a large increase in the number of gates in these new ASIC's compared with LHC designs and eventually to a significant increase in timing sensitivity across a chip and Single Event Upset sensitivity. This evolution in size, speed, and radiation requirements for HL-LHC designs led to the need for new verification techniques, which also implies the need for a larger design team and introduces the prospect of long redesign cycles for failed designs, which may take as long as a year \emph{crisis}.   
    CERN's electronics system design group (ESE) responded by proposing and forming a task force designated as CERN-HEP IC design  and platform services in microelectronics (CHIPS) to formally engage experts from CERN's ESE group with ASIC design teams experiencing problems to collaborate and broaden the range of experience available for problem solving and ensure that up to date  verification techniques are being employed. This expanded role for CERN's ESE group has provided additional important help for HL-LHC design groups beyond their traditional umbrella activities to facilitate uniform access to ASIC technology and design flows in addition to important IP design blocks that they  openly  share with the community. 
    An important exemplary way to effectively meet the need to  increased the size of the design team for highly complex ASIC design is the CERN RD53 group formed as a collaboration of multiple design groups to tackle the common issue of pixel designs in high radiation areas near the collision point of the inner trackers that are part of both the ATLAS and CMS detectors. In a recent submission only a  few masks were different between the ATLAS and CMS versions of the pixel chip.   It should be noted that a unified IP  management platform  with verification requirements for updating designs.

    \subsection{Low-cost High-Impact Online Infrastructure} 
    
    The emergence of the Web has been a significant help in the design of our present day detector systems and can be more intentionally  exploited to:   
\begin{itemize}
    \item Provide a broad Institutional Memory for:
        \begin{itemize} 
             \item  Links to archived HEP detector design documentation
             \item  Design specific optimization approaches and test bench development
             \item  Detector specific FAQ's: how element size or modularity or other design parameters were chosen 
             \end{itemize}
    \item Provide  self-guided training guides/examples for system/sub-system blocks.
    \item Provide hierarchical working areas for the creation and modeling (or modeling results) of  detector system and sub-systems 
            \begin{itemize} 
             \item             sub-system interface documentation 
             \item              sub-system designs
             \item Lists of development needs \& system dependencies
            \end{itemize} 
    \item Recommend PCB and ASIC  CAD design platforms, tools, design flow examples.
    \item Provide specific ASIC design examples and archive  FAQ's (avoiding NDA violating information)  
    \item Provide Link locations \& storage for community driven technical assistance.
\end{itemize} 
             
    \subsection{Organized Events and Training}  
          As the complexity of our detector systems have grown,  the number and type of specialists needed has increased dramatically.  To ensure community awareness of the state of the art as new ideas and techniques evolve over the long cycles of detector design and development it will be important to organize  specialty meetings that include both presentations and training as well as general instrumentation meetings.  Keeping these primarily in the US should minimize the cost.  In addition  recently DOE supported Traineeship opportunities (see 3.1 below) and multi-session training courses will be crucial for recruiting and maintaining our workforce. In addition, web-based self-paced tutorials developed in the workshops and schools described here will broaden the reach of the educational tools developed.
          
    The proposed activities are:
    \begin{itemize}   
     \item  Specialty workshops with training session options for a variety of topics such as:   FPGA design and resource allocation, ASIC transistor level analog design, digital library cell focused  design, Place \& route and  verification tools and techniques at the chip, module and system level. 
     
    \item {General meetings covering complete or nearly complete designs that integrate multiple specialties }  The Front End Electronics (FEE) workshop held on a bi-annual cycle for more than a decade  and the annual Topical Workshop on Electronics for Particle Physics are good examples and certainly should be continued. It should be noted that  presentations at these meetings have gravitated towards summary  "SOC" or System On a Chip descriptions that, out of necessity, have lost focus on intimate design details.  Thse SOC's  likely contain an analog front end, digitizing unit, data storage, and sparsification blocks with selective readout and internal calibration and test functions.  In the not so distant future our SOC's will likely benefit from a learning capability derived from artificial intelligence (AI) or machine learning (ML).
    
 %%    \items Traineeships
 %%       The DOE has recently introduced two Traineeship programs, one based at Labs and one at Universities with lab based specialty training. These offer access to Lab personnel, university Grad students, staff and interested faculty.  In the University program General Instrumentation courses are offered at the university level the traineeship program allow courses on detector technologies and electronics to be given by institutions with strong commitments to detector development.  This sharing of educational resources and teaching modules among US institutions of higher learning should be encouraged and broadened if possible.   ASIC design hands on  training is provided by labs with experienced staff providing mentoring.  This requires  that participants spend several months at their facility to avoid NDA issues.   Traineeship FOA's  are certainly a breakthrough effort on the part of DOE to renew and maintain the instrumentation workforce that will hopefully continue for decades. 
 %%       
     \item Instrumentation Specialist Degrees  for university physics students, until there are well understood instrumentation degrees or sub-majors, most grad students pursuing a career in HEP will participate in and hopefully lead some kind of Physics analysis towards a traditional PHD in the US.  They will get training in narrow lanes that concern the development of their degree institution's task at hand. A cultural shift needs to be promoted, namely the development of innovative instruments must be recognized a scientific research accomplishment suitable to represent a large portion of a doctoral dissertation in experimental particle physics. This is one of the goals pursued by the DPF CPAD panel. For students not lucky enough or aware of the Traineeship programs a  minimal level of training could be packaged in several one or two day topical workshops on Instrumentation that add some kind of independent study credits to their Experimental HEP degree. Curricular material can be developed to support dedicated courses to be incorporated in doctoral programs.
     In general, it is essential to give students a deeper understanding of the  workings of the detector with which the data they are analyzing are collected. 
\end{itemize}

   A summary of  Design tools (Table~\ref{table:tools}) and Foundries (Table~\ref{table:foundries}) used for ASIC design and fabrication are given in section 5   It can be seen from these lists that extensive familiarity with multiple tools is required for experts involved with today's HEP ASIC designs. In fact it is highly valuable, perhaps mandatory,  to have multiple experts with different focus areas on a design team.   
  For future projects it would be  useful  to set up a web based, open access,  repository to encourage the user community to document and share problems and solutions encountered with tools and technologies through FAQ's and design discussion threads as well as have experts available via special channels like Mattermost or Slack to help with problems encountered by HEP designs that  encounter the limits of the  validation tools or ASIC technology their design goals push them to explore. 
  We propose a support program at a national level to allow broad access to a common set of design tools and technologies to make it easier to share information and train new designers.  This would help limit training costs, improve the knowledge base and make it easier for the community to find well informed design reviewers.

 % More precisely, we need to address needs in these three different areas:
%\subsubsection{ASIC specific topics}

% \subsubsection{FPGA, hybrids, and peripheral electronics}

 \subsection{Field Programmable Gate Arrays (FPGAs)}
 
Field Programmable Gate Arrays (FPGAs) have proven to be a core component of all experiment data processing pipelines. These devices in their simplest form provide the glue logic that ties various commercial and custom components, including ASICs together in a larger system. At their most complex, FPGAs serve as powerful data processing units, providing data manipulation and data reduction functions in many large experiment data pipelines. FGPAs are extremely flexible and serve as data processing components at the Edge, at the intermediate data concentration level as well as in the back end data processing farm of large complex experiment data pipelines. While ASICs will continue to be a critical component of these data processing pipelines, FPGAs provide a degree of flexibility in their operation allowing data processing to be adjusted as the system is commissioned and lessons are learned about the unique needs of each deployed system.

Classical development cycles involve an RTL language such as Verilog or VHDL to be synthesized into logic using tools which are specific to the vendor of the FPGA family. In recent years higher level interfaces such as Matlab Model Composer and XIlinx HLS have gained momentum as tools which allows for more complex data and RF processing techniques to be deployed without low level knowledge of the FPGA fabrics. These tools have had mixed success as they do allow for easy deployment of these processing methodologies but don’t always allow for the most efficient use of the resources on the FPGA. While, the Matlab Model Composer flow is very powerful in its ability to deploy complex DSP algorithms to FPGAs, it also has some negatives in its ability to be properly revision-controlled and does not always provide for low-level tuning of resource usage on the target fabric. The HLS design flow has the advantage easy integration into a revision control system such as Git. HLS suffers in its infancy seeing significant changes in its design flow as the supporting vendor learns from their mistakes and makes improvements in the tool. 

This reality of the various state of these tools sets, and other less mature design flows from higher level languages, requires that the community using these tools continues to refine skills and adapt designs to new versions of the tool set and more efficient FPGA fabrics. ACAP engines provide a lot of potential as both a DSP and AI processing engines, its current use is still limited as the vendor continues to refine the tool sets which supports its use. Both Xilinx and Intel, the leading commercial vendors for FPGA fabrics, are very focused on the software co-processing market, leaving the areas of Edge processing at a loss, requiring the developers to be creative in ways to deploy advanced data processing at the edge.

Multiple laboratories in the community have demonstrated strengths in RF DSP and AI processing in FPGAs. The successful projects which have deployed these technologies now face the need to update their knowledge both in new FPGA vendor design flows and adapting their deployed solutions to newer technologies. The RFSOC platform from Xilinx has demonstrated itself as a game changer for RF stimulus and data processing platforms in its incorporation of high speed ADC and DACs with a large FPGA processing fabric. The inclusion of a powerful ARM-based processor with an accompanying real-time processing and graphics processing engine has provided a very powerful tool. The best use of this new platform is continuously being investigated and is already being deployed on some experiments. The continued investment into the institutional knowledge of these new technologies, the design flows which enable their use and the core libraries which can be reused across multiple experiments are key to the future of data processing at the edge, in the data processing pipeline and in the back end data processing engines. 

It is important that HEP  laboratories and universities collaborate together as much as possible with shared resources and knowledge related to the use of FPGAs in large experiments. Previous discussions about the future of next generation data acquisition systems have identified the need for a common shared library amongst the various institutions. The publishing of general purpose libraries needs to be encouraged along with encouraging community members to contribute back to libraries which have been provided in the public domain. A set of standard approaches to common interfaces such as streaming and memory access buses should be standardized as much as possible to ensure that libraries developed at different institutions can be integrated together without adding complex wrappers. When possible the interfaces supported by leading vendors should be adopted when possible. These interfaces and their supporting libraries should be lightweight, limiting the need for adoption of additional support software or other libraries as much as possible. Interfaces should be well defined and allow the mix and matching of solutions from various institutions.

%\subsubsection{Performance validation, radiation qualification, and test beams}

\section{Workforce}
Skilled workforce to undertake the broad instrumentation program discussed above is in short supply. In addition, the cyclic nature of detector constructions and upgrades make this problem particularly harsh in specific times when big upgrades happen simultaneously. An example is the "ASIC crisis" that occurred at CERN not long ago. A way of ameliorating this problem is to promote an R\&D effort on ASIC designs similar to the CERN RD program that allows a steadier work on advancing the new technologies of the future.

\subsection{DOE Traineeship Programs}

Custom, high-performance electronics systems and ASICs are ubiquitous in high-energy physics experiments. Engineers and physicists able to develop these systems are increasingly important to the success of new experiments as increases in experiment scope and precision drive continually more advanced electronic instrumentation. As described in the Basic Research Needs for High Energy Physics Detector Research and Development report, "such skilled personnel are highly sought after by industry and maintaining a pipeline of students and early career engineers and scientists with these skills is essential". However, the experiences at multiple institutes indicates such a pipeline is currently inadequate, as evinced by the difficulty experienced by HEP-related ASIC groups in attracting appropriate applicants due to fierce competition with industry.

To address this need to develop a training pipeline of skilled engineers focused on the needs of the high-energy physics community, the DOE funded several consortia. One consortium, focused on ASIC design, is called HEPIC (High-Energy Physics Integrated Circuits). Another is the TRAIN-MI program (High Energy Physics Instrumentation Traineeship in Michigan).

The HEPIC consortium is a group of several California Universities (Stanford University, the University of California at Davis, and the University of California at Santa Cruz) in association with National Laboratories with a HEP-related ASIC design activity. The purpose of HEPIC is to give top students exposure to the challenges and exciting work opportunities in the field of radiation detectors before they graduate. This early access is critical as HEP laboratories cannot compete with industry salaries for IC Designers. Therefore, the goal is to give students an exciting experience in HEP and "hook" them into the field early. Building strong personal networks with promising students will help retain these students in the HEP community as they graduate.

To give the students meaningful research experiences, HEPIC member labs have developed research activities appropriate for graduate students and provides fellowships for these students to experience cutting-edge design work in HEP-targeted ASICs before they reach industry. The intention is that some portion of these students will choose to stay within the HEP ASIC development community. The continual renewal of students is required due to the observed turnover as HEP IC designers migrate to more lucrative industry jobs. 

In addition to summer research experience in HEP provided by the National Labs, HEPIC also includes an educational component. ASIC design for HEP applications has several unique features and challenges that differentiate it from most industrial design specialities. A set of talks organized throughout the school year and a summer short course given at SLAC will provide HEP-specific context to the students' existing technical knowledge and will enhance retention. The virtual connectivity developed in the last couple of years due to the COVID19 restrictions can foster broader participation to some of these events. This is possible only with a standardization of some design tools. 

The TRAIN-MI instrumentation traineeship centers around a University-endorsed certification program and joins the existing Accelerator Science and Engineering Traineeship (ASET) at MSU to provide comprehensive training in two undersubscribed subfields of HEP. Students from high energy physics, nuclear physics and related fields, including electrical engineering, are able to take dedicated instrumentation courses to gain a rigorous, broad, and deep understanding of instrumentation. The program has the following key objectives:
\begin{itemize}
    \item Curriculum: Students will learn about detector technologies (scintillators, photon detectors, calorimeters, bolometers, time projection detectors), detector readout and electronics, including ASICs, radiation safety and materials physics. In addition, project management, lab safety and scientific communication will also be covered. Successful completion of the program will result in formal certification.  
    \item Research opportunities: Post coursework, students will be paired with research mentors at national laboratories or DOE-supported facilities where they will be able to apply their skills to a research project with a central instrumentation component. An important goal is to build connections and experiences at national laboratories that will enhance students’ ability to build careers in instrumentation.
\end{itemize}

The program curriculum is assessed by an independent panel to review course materials, evaluate student progress, and advise on future curriculum needs.  The program includes dedicated mentoring and career opportunities to connect students and their skills to employment opportunities and a community of HEP instrumentation.

\section{Tools}

To develop the custom electronic systems and ASICs needed to instrument new physics experiments, engineers use power Computer-Aided-Design, or CAD tools. Specialized CAD tools are used at all levels of electronic hardware development and in particular ASIC design depends on access to complex, costly CAD systems. A list of commonly used  CAD design \& verification  tools can be found in Table~\ref{table:tools}

\subsection{CAD Design Tools}

Electronic systems and custom ASICs are designed using specialized software tools called Computer-Aided-Design, or CAD tools. These tools are costly and represent a significant barrier to entry for research groups to participate in ASIC design in particular. While low-cost and open-source tools have made inroads in Electronic Systems and PCB design, is has not yet been the case in custom ASIC design. Therefore, institutions need to make significant investments in order to secure access to appropriate tools for custom ASIC design. Different tools are needed for analog schematic entry, simulation, physical layout and verification. Additional tools are required for the design of the digital section of mixed-signal ASICs. As design complexity grows and designs are targeted to more advanced CMOS nodes, the costs and number of tools required increases.

\subsection{Verification}

Hardware development, and in particular ASIC development, requires much higher initial quality than software. This is because it can be costly or inconvenient to update electronic systems or PCBs once manufactured, and it can be impossible to update ASICs to fix errors after fabrication. Therefore, design verification has become a key component of successful ASIC development. In fact, in industry, verification effort can exceed design effort in the development of a complex integrated circuit. 

An appropriate focus on design verification has not been possible in HEP-related ASIC developments. Resource constraints force design teams to focus on the design component of the development cycle and are unable to execute verification plans to the same extent as industry. This can sometimes lead to additional ASIC fabrication cycles and can stretch out development timelines. 
Another issue with verification is that it is, fundamentally, a skill set distinct from design. Best practice in industry is for a design to be verified using different personnel than the team who designed the chip. This is not possible, typically, in HEP-related developments due to small teams and limited resources.

As ASIC designs for HEP applications continue to increase in complexity and functionality \cite{Begel:2022wul}, the mismatch between verification needs and capacity will continue to grow and creative solutions will be required to deal with the growing complexity of HEP ASICs.

\subsection{Version Control}

Version Control is the managing and tracking of changes to designs. Version control allows rapid evaluation of new features, allows errors to be managed by "rolling back" to earlier versions of designs, and allows different versions of designs to be compared or reconfigured on the fly, saving time and development resources. Version control is ubiquitous in software development and is becoming increasing popular in hardware development. 

The most popular version control tool is the open-source software tool called git. Git is a lightweight version control system that also interfaces with web-based repositories (such as Github and Gitlab) that enhance ease-of-use and promote cross-site collaboration.

While Git can (and has been) used to track changes in hardware development, more specialized tools that interface directly with the CAD tools add significant value, particularly for analog designs. In the HEP ASIC design community, the version control system SOS by Cliosoft is widely used. It enables a design information repository to be mirrored across sites in an intelligent way where changes to the database at one site is propagated throughout the collaboration as needed. To update the design, a user "checks out" the design so it cannot be changed by others. When the design update is complete, the user can "check in" the updated design. When ready, other users can "update" their local repository, essentially re-synchronizing the data as seen by each collaborating sites. This enables progress to be made on multiple fronts simultaneously. 

Version control is a key technology for HEP-based collaborative developments. It is difficult to imagine how complex jointly designed ASICs (such as for the ATLAS and CMS experiments) could be developed without it. Therefore it is critical to maintain access to appropriate version control tools (such as Cliosoft SOS).

    \begin{table}[ht]
    \begin{center}
\begin{tabular}{ |c|l|l| } 
 \hline
 Platform & Tool & Description \\ 
 \hline
 Cadence &  &  \\ 
   & Virtuoso  & schematic capture and layout editor \\ 
   & NCverilog  & previous verilog simulator \\ 
   & Xcelium    &   current verilog simulator \\
   & Spectre & SPICE  analog simulation  \\ 
   & Genus & synthesis of gate-level logic from RTL  \\
   & Conformal & Logical equivalence checker \\
   & Joules & RTL-based power estimator \\
   & Liberate & Liberty (timing) file for PnR  \\
   & Innovus &  Floorplan; Place and Route \\
   & Voltus & Power simulations \\
   & Assura / PVS & one of several DRC/LVS tools \\
   & Tempus & timing signoff \\
   \hline
 Siemens  & & \\
   & Calibre & widely used  DRC/LVS \\
   & Eldo / Eldo Premier & SPICE analog simulator / fast SPICE \\
   & ADvancedMS & Mixed-mode (analog and digital) simulator \\
   & Questa (Modelsim) & RTL-based digital simulator \\
   & AFS & Fast SPICE \\
   & Tanner Suite & Low-cost design flow \\
   \hline
 Synopsys & & \\
   &  HSpice & SPICE simulations \\
   & FineSim &  fastSpice for full chip simulations\\
   & DC Compiler &  RTL synthesis\\
   & IC Compiler/IC Compiler II &  Floorplan, Place \& Route\\
   & PrimeTime & Timing signoff\\

 \hline
\end{tabular}
\end{center}
\caption{Tools commonly used in creation of ASICs for HEP}
\label{table:tools}
\end{table}

%\section{Foundries and test facilities}
\section{Commercial Foundries and Custom ASIC design}

\subsection{Commercial Foundries}

Custom ASICs for HEP application are almost exclusively fabricated using commercial foundries. See Table~\ref{table:foundries} for a  survey list of commercial foundries currently available for HEP designs below.  These foundries are companies that fabricate integrated circuits designed by independent design groups. Foundries are a key part of the "fabless" semicondutor model where many chip companies do not maintain their own fabrication capabilities but instead contract foundries to provide them with fabricated integrated circuits. Design groups in the DOE National Labs and in Universities can access the same foundries to make their custom integrated circuits. The foundries are typically accessed through Multi-Project Wafer (MPW) services offered by third-party brokers (such as IMEC/Europractice in Europe and MUSE in the USA). In an MPW, the broker aggregates unrelated designs from multiple customers into a single wafer submission, lowering the cost for each design group.

%  Lorenzo - move to section 5.2?
However, MPWs are not a viable solution for several specialized technologies. Moreover, some projects would also greatly benefit from having access to full wafers during the initial prototyping phase when the ASIC area is still limited to a few mm$^2$: for example, when advanced 3D interconnections or post-processing techniques are required. Limited access during the prototyping phase to a certain technology often results in the selection of a different one, often sacrificing performance for better accessibility. These limitations could be mitigated by facilitating collaborations between different Laboratories on the same technology, thus sharing the costs across multiple projects and increasing the bargaining power with the foundries. 

Design groups face several difficulties interacting with commercial foundries. First, these foundries (most notably Taiwan Semiconductor Manufacturing Company, or TSMC) have strict Non-Disclosure Agreements (NDAs) that contain clauses (such as indemnification and location of governing law) which are difficult for DOE-affiliated Laboratories to accept. Due to the mission-critical need of these agreements, and lack of satisfactory other options, the Labs typically make exceptions to policy to execute these agreements. This process can severely delay projects. It is not unusual to take a year or more to execute such agreements. 

As experimental needs continue to require ever increasing performance from the constituent ASICs, design groups are migrating to more advance semiconductor technology nodes. For example, much of the LHCb work largely used TSMC 130nm for Upgrade 1, and for Upgrade 2 the experiment is considering the use of finer-line CMOS technologies, such as 65nm or 28nm. The cost of semiconductor processing is a strong function of the process node and finer-line CMOS technologies require significantly more designer effort to manage the increased complexities. In addition, the foundries are much tighter in providing access to finer line CMOS technologies. 
%J. Fast
Additionally, porting designs between foundries/processes at the same node size is non-trivial so having assured access to a specific process at a specific foundry over the lifetime of a project is essential.

Maintaining and expanding access to commercial semiconductor foundries is crucial to maintaining a custom ASIC development capability in the US HEP community. Successfully executing HEP-related ASIC projects is becoming more difficult due to the increased design and fabrication costs of more advanced CMOS nodes, coupled with the increased difficulty of getting access and executing joint-NDAs, which are critical for cross-group collaboration.

%From J. Fast
OHEP is not alone in its concerns about assured access to commercial foundries. DOE-NNSA, DOD, NGA, NSA, NASA and others also rely on commercial products and often have much longer time horizons for needing stable supply chains. The NNSA and related DOD applications are perhaps the most relevant to OHEP/ONP as they have stringent strategic radiation hardness requirements. In addition, the NNSA and others have security and trust constraints that require domestic supply chains. U.S. Government (USG) foundries such as the MESA facility at Sandia National Lab and the MIT Lincoln Lab foundry provide assured access, security and trust\footnote{USG agencies utilize the Trusted Access Program Office (TAPO) of the Defense Microelectronics Activity (https://www.dmea.osd.mil/) to certify trusted suppliers.}  for the most mission critical applications at modes from 350 nm down to 90 nm. However, it is not currently feasible for the USG to operate a dedicated foundry at advanced ($<$90 nm) node size. Alternative solutions such as heterogeneous integration of commercial advanced node parts with in-house trusted components are being pursued to enable utilization of advanced nodes. The defense and intelligence community programs have significant leverage to shape U.S. policy towards domestic supply chain, though even combined lack sufficient funding to shape commercial enterprise planning for state of the art foundries. However, they do enter into long-term strategic partnerships with domestic foundries and can provide significant funding through Title III to support state of the practice foundries. In this regard, the science community can benefit from an understanding of what nodes and processes/foundries are likely to have long-term USG support.

Perhaps the most challenging aspect of managing development cycles is the rapid pace at which industry evolves - for better or worse. As a current example, Intel has recently launched Intel Foundry Services, essentially opening access to their cutting edge foundries. As part of that strategy they are purchasing Tower Semiconductor - the foundry at which CERN is developing the ITS3 65 nm MAPS devices. It remains to be seen whether this development is beneficial or detrimental to HEP.
%end J. Fast addition

A survey of  HEP labs and universities  produced the list of  ASIC foundries  in Table~\ref{table:foundries}  in use or under consideration. 
    \begin{table}[h!]
 \begin{center}
\begin{tabular}{ |c|l|l|c|c|l| } 
 \hline
Foundry &	Broker &	Processes &	MPW & Dig Cells   &	Technology  \\ 
 \hline
TSMC &	IMEC &	250nm &	yes & yes &	CMOS \\
    	& and &  	180nm &	yes &	yes &	CMOS \\
		& MUSE &  	130nm &	yes & yes &		CMOS \\
		&  &  	90nm  &	yes & yes &		CMOS \\
		&  &  	65nm  &	yes & yes &		CMOS \\
		&  &  	40nm  &	yes & yes &		CMOS \\
		&  &  	28nm  &	yes & yes &		CMOS  \\
		 \hline

Global Foundries & Direct &	22nn FDX &	yes &	yes &	CMOS \\
		&  &  	55nm &	yes	& &	CMOS \\
		&  &  	130nm &	yes &	yes &	CMOS \\ 
		\hline

  Intel &	Mosis &	22nm  	 &	yes &	yes & FFL	\\ 
  \hline
					
TowerJazz  (sub-div. Intel)  &	Direct & 350nm  & No &	yes	 & BiCMOS \\
 &	 & 180nm  &yes &	yes &	BiCMOS \\
 &	 & 65nm  &yes &	yes &	BiCMOS \\
 \hline   					     
 Skywater (OA PDKs) & Direct & 130nm  &	yes &	yes &	CMOS \\
	&	& 90nm   &	yes & yes &	CMOS \\
\hline
\vspace{1pt}
 Specialty Foundries & & & & &  \\	
 \hline
Xfab(www.xfab.com) 	& Direct & HV 1um,  0.6um &	yes &  &	\\	
& &	HV 350nm  & & & \\
& &	HV 180nm &	yes	 & & \\	
& &	1um &	yes  & MLM	 &	CMOS\&MEMS  \\
& &	.8um &	yes  & MLM	 &	Sensors RF  \\
& &	.6um &	yes  & MLM	 &	SiC \& GaN \\
& &	350nm &	yes  & MLM	 &	3D integration \\
& &	180nm &	yes  & MLM	 &	MicroFluids \\
			 	\hline	
Lfoundry (www.lfoundry.com) &	Direct &	150nm &	yes &	yes & CMOS \\
	& & 	110nm	 & 	& &  HV LDMOS  80-200V \\
	& & 		 & 	& &	Opto \\
	& &		 & 	& &	RF \\
   
 \hline
\end{tabular}
\end{center}
\caption{ ASIC foundries  for in use or  under consideration for  HEP projects}
\label{table:foundries}
\end{table}

\subsection{Commercial Custom ASIC Design}
As noted above, successive generations of data acquisition ASICs designed to meet the needs of HEP experimentalists require ready access to the latest tools and techniques, as well as the development, curation, and leverage for the needs of the future of the significant body of institutional knowledge that has evolved over the years. However, although essentially all major HEP detectors require significant innovation to meet the challenges of the moment, many of the innovations of previous HEP experiments, as well as those from fields outside of HEP, can also be just as relevant to experimental projects under development and require not ground-breaking R\&D but the effective leveraging of known principles to meet current needs. Similarly, as mentioned in the previous section, significant barriers exist in developing HEP-related ASIC projects in advanced technology nodes because of the technical complexity and cost of the required design tools, as well as requirements for joint NDA and IP agreements which are often imposed by academic institutions on any R\&D projects in which their workforce engages. Structural problems such as these can introduce significant inefficiencies across entire programmatic funding lines if multiple isolated design groups are each, in essence, "reinventing the wheel."

One solution to this lies in the collaboration of HEP detector electronics groups, within both academia and national labs, as well as the cross-cutting groups found within essentially all detector collaborations, with commercial custom ASIC design firms that specialize in the design, fabrication and long-term operational support of state-of-the-art data acquisition ASICs specifically optimized for given HEP applications. Custom ASIC design firms can bring very significant value and capabilities to the development of HEP detector electronics by offering substantial preexisting facilities (e.g. powerful commercial tools, licenses, etc.), expertise and experience in ASIC design and fab in many technology nodes, including advanced ones, as well as e.g. in the design of optimized analog and mixed input stages and  hybrid detector designs, along with implementation of on-chip support for the advanced capabilities which will be required to reduce to a manageable (and affordable) level the huge raw data rates and data set sizes expected in many next-generation experiments such as e.g. streaming trigger-less DAQ, feature extraction and zero suppression, automated calibrations, etc. as well as many other complex capabilities implemented in firmware such as e.g. event building, edge ML/AI inference and other advanced data flow functionality.

Programmatic mechanisms such as the Small Business Innovation Research (SBIR) and Small Business Technology Transfer (STTR) programs enable the most capable domestic small businesses to engage in Federally supported R/R\&D to synergistically develop with HEP experimentalists, as well as scientists engaged in research supported by DOE and NSF and across more than a dozen other Federal agencies, advanced technologies that have the potential for commercialization. And conversely, it allows HEP to efficiently and effectively leverage for their own needs advanced technologies which are developed in public-private collaboration across the entire breadth of the vast publicly funded R\&D enterprise. Such public-private partnerships, both small and large, are the foundation for diverse segments of the U.S. and global economy, and have historically allowed the often paradigm-changing benefits of R/R\&D to flow from the laboratories of publicly funded research institutions into the wider society for the benefit of all. The continuing evolution and increasing support for mechanisms facilitating such public-private partnerships such as e.g. SBIR/STTR, Cooperative Research and Development Agreements, etc. will serve the dual purpose of better enabling HEP physicists to construct the experimental instruments of the future while making the fruits of their labors on behalf of HEP more directly accessible to scientists and researchers across the entirety of the public and private sectors and, ultimately, into the lives of us all.

% \subsection{Research Friendly (Specialty) Foundry services }

% \subsection{Test facilities}

\section{Open Source}
An interesting trend in hardware development has been the introduction of Open Source ideas borrowed from the software development world. One important example is the introduction of the RISC-V instruction set. This is a new microprocessor instruction set architecture (ISA) that is free of patents, licence requirements, or other encumbrances. The RISC-V ISA has unleashed a significant amount of design creativity and has helped drive Open-Source efforts in other areas of hardware development.

\subsection{CAD Tools}

CAD tools from various University groups have been available for many years. Venerable tools such as Magic (for mask layout editing) have been used by generations of students in coursework and research. However, as integrated circuits get more complex, there has been a need for integrated design flows to improve design productivity. It has been difficult for open-source tools to keep up with this trend due to the significant investments needed to ensure various tools can work together to effectively develop high-performance integrated circuits. For this reason, design groups in the HEP community have relied on costly commercial tools. This has created a significant barrier to entry for new groups to address the needs and challenges of electronics for HEP.

Initiatives such as DARPA's POSH (Posh Open-Source Hardware) and IDEA (Intelligent Design of Electronic Assets) programs have helped focus attention on these and the need for high performance open-source hardware development flows. Frameworks such as OpenRoad are making significant strides in increasing design productivity at low or no cost. 

\subsection{Foundry Processes (PDKs)}

In addition to CAD tools, another important barrier to entry is difficulty in accessing semiconductor foundry confidential information. An innovative approach to this issue is the concept of an Open-Source PDK (Process Design Kit). As part of an initiative driven by Google, SkyWater Technology has open-sourced their SKY130 130 nm CMOS process. This means anyone can download the technical manuals, model files, and the like without entering into an NDA. 

Open-source PDKs in conjunction with open-source CAD tools hold the potential to increase HEP-related development activities by lowering costs and barriers to entry, and also offer intriguing new opportunities, such as rapid design-space exploration, which are not practical under the current cost model.

\section{Collaborative R\&D efforts for sustained electronics and ASIC capabilities}

\subsection{CERN sponsored RD groups}

% Lorenzo: need to expand more.
A recent example of collaborative R\&D effort lead by CERN is the WP 1.2, which includes several European and US-based research institutes. With the ALICE ITS3 upgrade as main driver of such project, several groups have gained access to the TowerJazz 65 nm imaging technology for the development of a new generation of Monolithic Active Pixel Sensors (MAPS), sharing knowledge about the technology and design blocks. Another example is the "28 nm Forum", where more than 100 participants are sharing knowledge about the TSMC 28 nm technology, a candidate for the next generation of detectors at HL-LHC.

\subsection{Opportunities for new RD groups in the US}
There are several recent developments in the broad area of instrumentation of particle detectors that could benefit from the kind of structured long-term planning, vision, and support embodied by the CERN RD Collaboration model. One potential technology that could benefit from a coordinated US-based collaboration model would be scalable pixelated detector systems. This is a new development in the readout of Liquid Argon Time Projection Chambers (LArTPCs) that provide unambiguous 3D imaging of the detector but must meet stringent requirements on noise, power, and reliability in a cryogenic environment. The first generation LArPix ASIC demonstrated the feasibility and promise of such an approach. There are many areas in which the systems need development to realize the goal of scalable detector readout systems. For example, future R\&D activities will pursue finer detector granularity, improvements in embedded detector logic, increased system reliability, and advances in commercial mass production. Other potential areas of research include higher bandwidth detector systems, adaptable readout logic, and large-area photodetection to extend the concept to light detection.

Similar to how CERN organizes RD collaborations, the DOE, though the national laboratories, could provide a similar shared infrastructure and sustained, shared support among a large number of university and laboratory partners. The scalable pixelated detector R\&D proposed here could serve as a test case for such a model within the US. The CERN RD collaboration have been successful at sharing expertise, providing common tools and process, coordinating activities, and disseminating the knowledge and technical expertise that underlie the current generation of experiments. The US detector development community could benefit greatly from such a shared deveopment model.

\section{Conclusions}

To continue the rate of progress in US-based HEP research, a highly trained, experienced, and responsive workforce is needed to supply the custom ASICs that are critical to the success of advanced detectors. Improved access to training, retention of a skilled work force, and preservation of the vast knowledge accumulated in decades of research are necessary elements of this program. Attention to the development of diverse work force is an important consideration. A challenge is the identifications of the resources to support the educational and technical infrastructure to implement this program. 

A consolidation and broader availability of the design tools and access to foundries to develop the microelectronics components needed for future particle physics experiments is another pillar of the ambitious detector development program envisage for the near future.

To simultaneously develop the advanced technologies needed for future detectors, and to ensure we maintain a robust, highly trained ASIC and electronics development workforce, we propose that the US HEP community undertake long-term, collaborative, directed R\&D projects comparable to the RD program led by CERN. This would enhance retention of design talent by filling in the gaps of ASIC design between large projects, and help to shorten detector and electronics infrastructure development time, as key technologies could be ready when they are needed by the projects.

%%%%%%%%%%%%%%%%%%%%%%%%%%%%%%%%%%%%%%%%%%%%%%%%%%%%%%%%%%%%%%%%%%%%%%%%%
% example figure

%\begin{figure}
%\begin{center}
%\includegraphics[width=0.40\hsize]{xxx}
%\end{center}
%\caption{xxx}
%\label{fig:xxx}
%\end{figure}

%%%%%%%%%%%%%%%%%%%%%%%%%%%%%%%%%%%%%%%%%%%%%%%%%%%%%%%%%%%%%%%%%%%%%%%%%%%

%%%%%%%%%%%%%%%%%%%%%%%%%%%%%%%%%%%%%%%%%%

%  If you would like to use BibTEX for the bibliography, please feel free to do so.  It is not required.

%  To use BibTeX,

%    1.  uncomment the following two lines, 
%    2.  comment out everything below from  \begin{thebibliography}{99}   to \end{thebibliography).
%    3.  create the file  myreferences.bib, and process this file in the usual way
\addcontentsline{toc}{section}{References}
%\setboolean{inbibliography}{true}
\bibliographystyle{JHEP}
\bibliography{references}  % file %myreferences.bib

%%%%%%%%%%%%%%%%%%%%%%%%%%%%%%%%%%%%%%%%%
%example bibliography

%\begin{thebibliography}{99}

%  this is a vanilla LaTeX bibliography.  It is also fine to use
%  bibTeX, but please be sure that bibTeX does not mangle your
%  citations

%\bibitem{}

%\end{thebibliography}

\end{document}